\documentclass[aps,prl,preprint,groupedaddress]{revtex4-1}
\usepackage{bbm}
\usepackage{graphicx}
\usepackage{nicefrac}
\usepackage{amsmath}								
\usepackage{amsfonts}								
\usepackage{amssymb}								
\usepackage{amsthm}								
\usepackage{float}

\begin{document}


\title{\textbf{The Quantization of the Rabi Hamiltonian}}
\author{Eva R. J. Vandaele, Athanasios G. Arvanitidis, Arnout Ceulemans}
\affiliation{{Department of Chemistry, Katholieke Universiteit Leuven, \\ Celestijnenlaan 200F, B-3001 Leuven, Begium, \\ arnout.ceulemans@chem.kuleuven.be}}

\date{\today}

\begin{abstract}
The Bargmann-Fock representation of the Rabi Hamiltonian is expressed by a system of two coupled first-order differential equations in the complex field, which may be rewritten in a canonical form under the Birkhoff transformation. The
transformation gives rise to leapfrog recurrence relations, from which the eigenvalues and eigenvectors could be obtained. The interesting feature of this approach is that it generates integer quantum numbers, which relate the solutions to the Juddian baselines. The relationship with Braak's integrability claim [PRL \textbf{107}, 100401 (2011)] is discussed.
\end{abstract}
\maketitle
\section{Introduction}
The Rabi Hamiltonian describes the coupling of a two-level fermion system with a single bosonic mode. In spite of its extreme simplicity, it shows up in an incredible range of applications, and moreover continues to stimulate further theoretical work.
In the focus point of much of the recent theoretical developments is the integrability claim by Braak \cite{Braak}. Braak derived a transcendental function, the zeros of which correspond to the energy spectrum of the Rabi Model. Each level is characterized by two quantum numbers, consisting of a two-valued parity label distinguishing symmetric and anti-symmetric states, and an integer, counting the nodes of the transcendental function.
States with the same parity do not show level crossings. For this reason the two quantum numbers together provide a unique labeling of each individual state of the system, which therefore is said to be quantum integrable.
So far the integer quantum number remains a mere counting number, with no apparent relation to the nature of the corresponding quantum state. In the present study we consider a further transformation of the Rabi Hamiltonian to a canonical form, which gives rise to a quantization condition expressed in integer numbers.
\section{The model system}
The model system consists of two fermion states, coupled to a single harmonic oscillator. It is a two-parameter system: the fermion level splitting is parametrized as $2\Delta$ and the linear vibronic coupling parameter is represented by $g$. The oscillator quantum $\hbar \omega$ is taken as the unit of energy.
The corresponding adiabatic potential energy surface is represented by a two-well potential with an avoided crossing, as shown in Fig. 1.

\begin{figure}
        \includegraphics*[width=10.8cm]{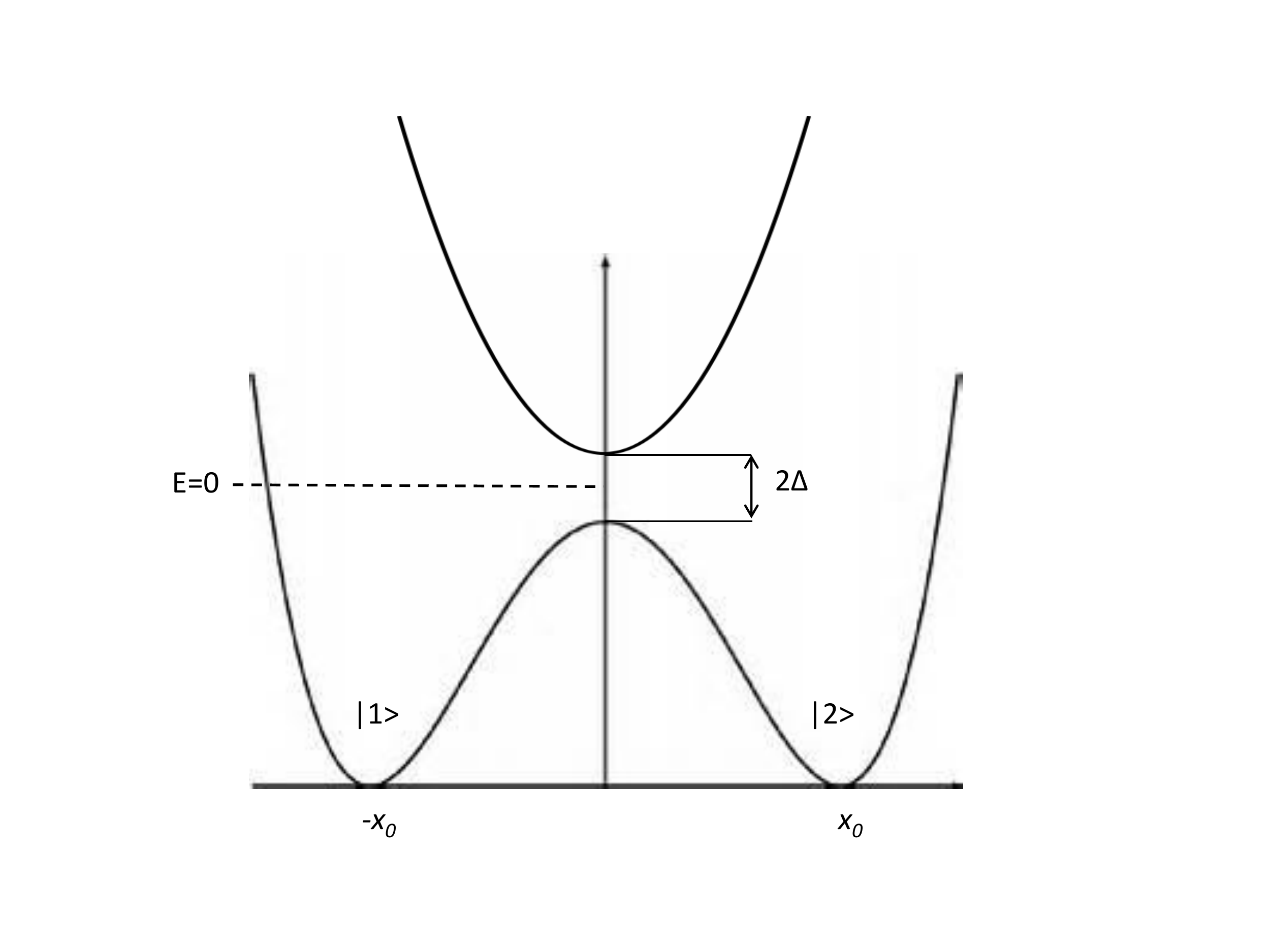}
        \caption {Two-well potential energy surface. The displacement $x_0$ is $\sqrt{2}g$.}
        \label{fig1}
\end{figure}

The potential surface has a reflection symmetry in the displacement coordinate. This property plays an important role in the quantization of the Rabi Hamiltonian. The localized electronic states in the separate wells will be denoted as 
$|1\rangle$ and $|2\rangle$.
The level splitting acts as a constant off-diagonal coupling term between the localized states:
\begin{equation}
H_{12} = \Delta \left\{ \left| 1 \rangle \langle 2 \right|+\left| 2 \rangle \langle 1 \right|\right\}
\end{equation}
Further developments have been considered where a bias between the two wells is introduced, but we will limit ourselves here to the simplest symmetric case.  
The Hamiltonian can be written in a matrix form, acting 
in the space $\left( \begin{array}{*{1}{c}}
			   |1\rangle \\
			   |2\rangle
			    \end{array} \right)$,
\begin{equation}
\mathcal{H} - \nicefrac{1}{2} \mathbb{I} =
\left( {\begin{array}{*{20}{c}}
			   a^\dagger a +g\left( a^\dagger +a \right) & \Delta  \\
			   \Delta & a^\dagger a -g\left( a^\dagger +a \right)
			\end{array} } \right)
\end{equation}

The corresponding vibronic wavefunction is a combination of the fermion states with 
coefficients that are functions of the boson excitations.
The solution of the corresponding Schr\"odinger equation can be obtained by expanding the coefficients in the boson space. The further treatment depends on the choice of the oscillator basis set: either centered in the coordinate origin or displaced towards the minimum of the well.
\subsection{Central basis set}
A straightforward basis sets consists of the unperturbed oscillator functions, centered in the coordinate origin:
\begin{equation}
|n\rangle = \frac{1}{\sqrt{n!}}\left(a^\dagger\right)^n |0\rangle
\end{equation}
The reflection symmetry, $\hat{\sigma}$, will affect both the boson and the fermion parts as follows:
\begin{eqnarray}
\hat{\sigma} (a^\dagger) &=& - \, a^\dagger \nonumber \\
\hat{\sigma} (a) &=& - \, a\nonumber \\
\hat{\sigma} |1\rangle &=& |2\rangle \nonumber \\
\hat{\sigma} |2\rangle &=& |1\rangle
\end{eqnarray}
Since the reflection plane is a binary symmetry element it has two representations: a symmetric and an anti-symmetric one, hence one has:
\begin{equation}
\hat{\sigma} |\Psi\rangle = \pm |\Psi\rangle
\end{equation}
As a result the Ansatz for the symmetric and anti-symmetric eigenfunctions becomes:
\begin{eqnarray}
|\Psi_+\rangle &=& \sum\limits_{n=0}^\infty c_{n} \,\frac{1}{\sqrt{n!}}\left(a^\dagger\right)^n |0\rangle \times
\frac{1}{\sqrt{2}} \left\{ |1\rangle +
(-1)^n |2\rangle\right\} \nonumber \\
|\Psi_-\rangle &=& \sum\limits_{n=0}^\infty c_{n} \,\frac{1}{\sqrt{n!}}\left(a^\dagger\right)^n |0\rangle \times
\frac{1}{\sqrt{2}}  \left\{ |1\rangle -
(-1)^n |2\rangle\right\}
\end{eqnarray}
The corresponding Hamiltonian matrices are:
\begin{equation}
\mathbb{H}_\pm -  \nicefrac{1}{2} \, \mathbb{I}=
\left( \begin{array}{*{30}{c}}
\pm \Delta & g & 0 & 0 & 0 & \cdots \\
g & 1\mp \Delta & \sqrt{2} g & 0 & 0 & \cdots \\
0 & \sqrt{2} g & 2 \pm \Delta & \sqrt{3} g & 0 & \cdots \\
0 & 0 & \sqrt{3} g & 3\mp \Delta & \sqrt{4} g & \cdots \\
0 & 0 & 0 & \sqrt{4} g & 4 \pm \Delta & \cdots \\
\cdots & \cdots & \cdots & \cdots & \cdots &\cdots 
\end{array} \right)
\end{equation}
The roots can then be obtained by straightforward diagonalization of the truncated matrices. Convergence is rather slow since the increase of the diagonal elements is offset by the simultaneous increase of the off-diagonal elements. Note that a sign change of $\Delta$ leads to a switch of symmetric and anti-symmetric solutions.
\subsection{Displaced oscillator states}
An alternative basis set makes use of displaced oscillator states which localize the oscillator in one of the wells. This is the basis set used by Braak in the framework of the Bargmann-Fock mapping. Later on Chen et al. rederived Braak's results in the Schr\"odinger representation \cite{Chen}. In the latter setting $|n\rangle_A$ denotes the oscillator state displaced to the well on the left, and $\hat{D}(g)$ is the displacement operator:
\begin{eqnarray}
A^\dagger &=& a^\dagger + g \nonumber \\
|n\rangle_A &=& \frac{1}{\sqrt{n!}}\left(A^\dagger\right)|0\rangle_A  \nonumber \\
|0\rangle_A&=& \hat{D}(g) |0\rangle = \exp(-\frac{g^2}{2})\exp(-ga^\dagger) |0\rangle
\end{eqnarray} 
In the formalism of Chen et al. \cite{Chen} the wavefunction is expressed in the displaced oscillator states as:
\begin{equation}
|\Psi\rangle_A = \sum\limits_{n=0}^\infty \left( \sqrt{n!} \,e_n |n\rangle_A \left|1\right\rangle +  \sqrt{n!}\, f_n |n\rangle_A \left|2\right\rangle \right)
\end{equation}
The action of the Hamiltonian on this wavefunction finally leads to the following 
defining relations for the coefficients:
\begin{eqnarray}
0 &=& e_m +\Delta \frac{f_m}{m -g^2-E} \nonumber \\
0&=&\Delta \, e_m + \left(m+3g^2-E\right) f_m - 2g(m+1) f_{m+1} 
-2g (m-1) f_{m-1}
\end{eqnarray}
from which one may derive the three-term recurrence in the $f$-coefficients:
\begin{equation}
f_{n+1}- \frac{1}{n+1} \Omega_{n} f_{n} + \frac{1}{n+1} f_{n-1}=0
\end{equation}
with:
\begin{equation}
\Omega_{n} = \frac{1}{2g} \left( n+3g^2 - E - \frac{\Delta^2}{n-g^2-E} \right)
\end{equation}
In this treatment the zero-point energy of $\nicefrac{1}{2}\hbar\omega$ is incorporated into the $E$ variable. The series starts off at $f_0$, which we can fix to unity; then one has:
\begin{eqnarray}
f_0 &\equiv & 1 \nonumber \\
f_1 &=& \Omega_0 \nonumber \\
f_2 &=& \frac{1}{2} \left( \Omega_0 \Omega_1 -1 \right) 
\; \:
etc.
\end{eqnarray}
The recurrence may be rewritten as:
\begin{equation}
\frac{f_{n+1}}{f_n} - \frac{1}{n+1} \Omega_{n}  + \frac{1}{n+1} \frac{f_{n-1}}{f_n}=0
\end{equation}
Let $t$ be defined as the limiting value of the ratio between consecutive coefficients:
\begin{equation}
t=\lim_{n\rightarrow \infty} \frac{f_{n+1}}{f_n} 
\end{equation}
The roots of the corresponding characteristic Poincar\'e polynomial \cite{Gautschi} are then given by:
\begin{eqnarray}
t_1 &=& 0 \nonumber \\
t_2 &=& \frac{1}{2g}
\end{eqnarray}
The zero root corresponds to the {\em minimal} solution, which is square integrable, while the upper root is the {\em  dominant} solution which converges to a finite non-zero value. The solution of the recurrence relations should be aiming at finding the minimal solution which is the only one that satisfies the quantization condition.
Moroz has argued that it is possible to obtain the eigenenergies by truncating the recurrence for sufficiently high $n$, and requiring the series to start at $f_0=1$ \cite{Moroz, Moroz2014}.The difference between the fixed value from $f_0$ and the initial value of the truncated recurrence relation is then Moroz' transcendental function, $F_0$, which reproduces the eigenenergies. This function is shown in Figure 2.
\begin{figure*}
        \includegraphics*[width=\columnwidth]{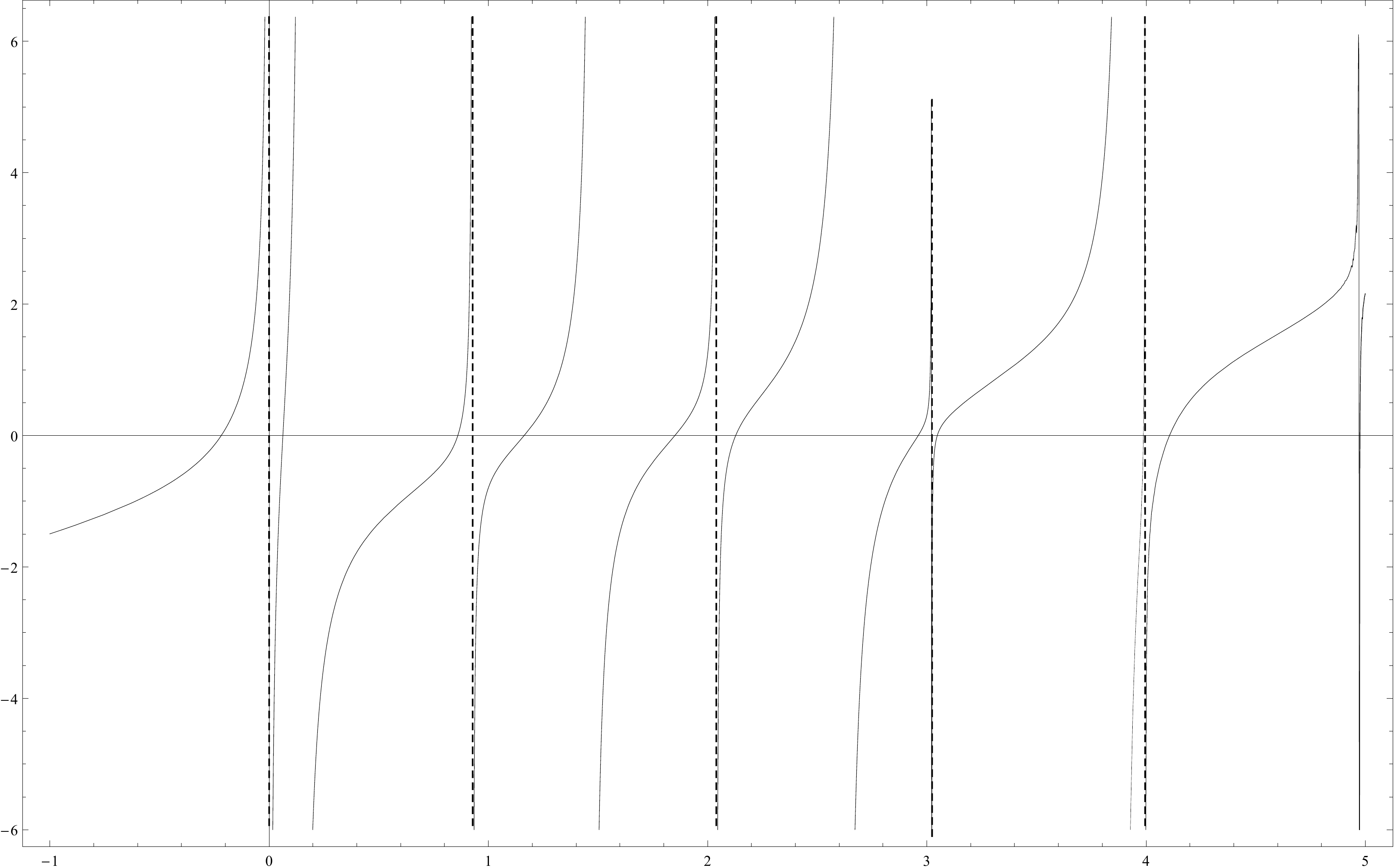}
        \caption{Plot of Moroz' $F_0(E+g^2)$ in the interval [-1,5] for $g=0.7$ and $\Delta=0.4$.}
        \label{fig2}
\end{figure*}

There is no guarantee though that this truncation will always lead to the minimal solution. Moreover so far no use was made of the reflection symmetry, which commutes with the Hamiltonian. As argued by Braak \cite{Braak2013} both aspects are related since the reflection symmetry can be used to construct a new transcendental function which will always lead to the minimal solution.
Applying $\hat{\sigma}$ to the oscillator functions maps the displacement from one well to the other. Let the oscillator basis functions in the well to the right be represented as 
$|n\rangle_B$. One then has:
\begin{equation}
\hat{\sigma} |n \rangle_A = (-1)^n |n\rangle_B
\end{equation} 
The reflection of $|\Psi\rangle_A$ then is given by:
\begin{equation}
|\Psi \rangle_B =
\sum\limits_{n=0}^\infty \left( \sqrt{n!} \,(-1)^n f_n |n\rangle_B \left|1\right\rangle +  \sqrt{n!}\,(-1)^n e_n |n\rangle_B \left|2\right\rangle \right)
\end{equation}
Since the exact wavefunction must have reflection symmetry and has to be single-valued,
one has:
\begin{equation}
|\Psi \rangle_A = \pm |\Psi \rangle_B
\end{equation}
Repeating now the derivation of the recurrence relations finally leads to a new transcendental function, $G_\pm(E)$, originally proposed by Braak \cite{Braak}, that has the eigenvalues as roots:
\begin{equation}
G_\pm(E)= \sum\limits_{n=0}^\infty g^n f_n(E) \left[ 1\pm \frac{\Delta}{n-g^2-E}
\right]
\end{equation}
Here the $\pm$ sign refers to symmetric versus anti-symmetric solutions respectively.
This function is displayed in Fig 3.
\begin{figure}
        \includegraphics*[width=\columnwidth]{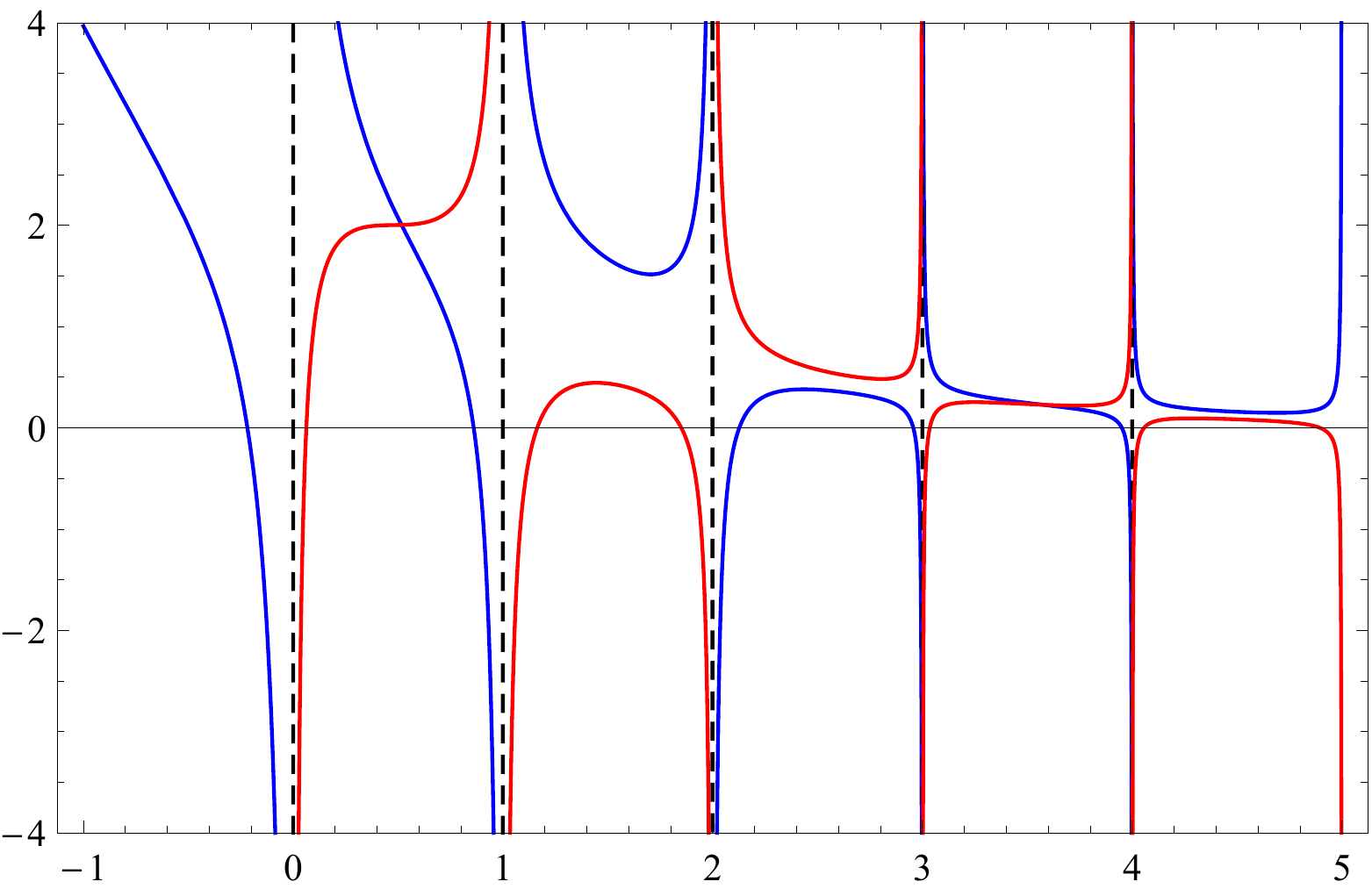}
        \caption{Braak's transcendental function $G_\pm(E+g^2)$. Red and blue lines represent the symmetric and anti-symmetric roots resp. in the interval [-1,5] for $g=0.7$ and $\Delta=0.4$ }
        \label{fig3}
\end{figure}

\section{The Birkhoff transformation}
We now continue the treatment in the Bargmann-Fock mapping. In this mapping \cite{Bargmann} the creation and annihilation operators are replaced by a complex variable $z$, and its derivative respectively: $a^\dagger \rightarrow z$, $a
\rightarrow \frac{d}{dz}$. The requirement that both operators remain each others adjoint is taken into account by defining the inner product of two functions as follows:
\begin{equation}
\langle f|g \rangle = \frac{1}{\pi} \int {\overline{f(z)}} g(z) \exp(-\bar{z}z) dxdy,
\;\; z=x+iy
\end{equation}
As a result the Schr\"{o}dinger equation is transformed into a set of two coupled first-order differential equations in $z$.
\begin{eqnarray}
\frac{d}{dz} f_1 &=& \frac{E-gz}{z+g} f_1 - \frac{\Delta}{z+g} f_2 \nonumber \\
\frac{d}{dz} f_2 &=& - \frac{\Delta}{z-g} f_1+\frac{E+gz}{z-g} f_2 
\end{eqnarray}
Note that the reflection symmetry, $(z) \leftrightarrow (-z)$, for this set of equations is preserved:
\begin{equation}
\mathbf{f}(z) = \left( \begin{array}{*{1}{c}} f_1(z) \\ f_2(z) \end{array}  \right)
=\pm \left( \begin{array}{*{1}{c}} f_2(-z) \\ f_1(-z) \end{array}  \right)
\end{equation}
The set of equations has two finite singular points at $z=\pm g$. Following an earlier treatment \cite{Ceulemans1996} we now apply the Birkhoff transformation \cite{Birkhoff}, which provides a method to remove these singular points to infinity.  Let us first rewrite the equations in a more general way as:
\begin{eqnarray}
\frac{d}{dz}f_1 =  p_{11}(z) f_1 +p_{12}(z) f_2 \nonumber \\
\frac{d}{dz}f_2 =  p_{21}(z) f_1 +p_{22}(z) f_2
\end{eqnarray}
or, 
\begin{equation}
\frac{d}{dz} \mathbf{f} = \mathbbm{p} \,\mathbf{f}
\end{equation}
Outside the circle $|z|=g$ the $p_{ij}$-coefficients can be expanded in a Laurent series
\begin{equation}
p_{ij} = \sum\limits_{k=-\infty}^q p_{ij}^{(k)} z^k, \: p_{ij}^{(k)} \in \mathbf{C}
\end{equation}
Here $q+1$ is the rank of the singular point at infinity. For the Rabi Hamiltonian the rank is equal to 1, hence $q=0$. The corresponding series expansions are as follows:
\begin{eqnarray}
p_{11}(z) &=& \frac{E-gz}{z+g} \nonumber \\
&=& -g + \frac{E+g^2}{z} -\frac{g(E+g^2)}{z^2}
+\frac{g^2(E+g^2)}{z^3}-\frac{g^3(E+g^2)}{z^4} + O(\frac{1}{z^5}) \nonumber \\
p_{12}(z) &=& -\frac{\Delta}{z+g}=-\frac{\Delta}{z} + \frac{g\Delta}{z^2}  
-\frac{g^2\Delta}{z^3} +\frac{g^3\Delta}{z^4} - O(\frac{1}{z^5}) \nonumber \\
p_{21}(z) &=& -\frac{\Delta}{z-g}=-p_{12}(-z) \nonumber \\
p_{22}(z) &=& \frac{E+gz}{z-g}=-p_{11}(-z)
\end{eqnarray} 
Now we assume a linear transformation of the form:
\begin{equation}
\mathbf{f} = \mathbbm{a} \mathbf{F}
\end{equation}
where the transformation coefficients $a_{ij}(z)$ are analytic at infinity and reduce at infinity to the unit matrix:
\begin{equation}
a_{ij}(z) =\sum\limits_{k=0}^\infty \frac{a_{ij}^{(k)}}{z^k} \: , \:\: a_{ij}^{(k)} \in \mathbf{C}
\end{equation}
This matrix transformation contains all the finite singularities of the initial system.
In view of the symmetry of this matrix we adopt a simplified notation as:
\begin{eqnarray}
a_{11}(z) &=& \sum\limits_{k=0}^\infty \frac{a_{k}}{z^k} \nonumber \\
a_{12}(z) &=& \sum\limits_{k=0}^\infty \frac{b_{k}}{z^k} \nonumber \\
a_{21}(z) &=& a_{12}(-z) \nonumber \\
a_{22}(z) &=& a_{11} (-z)
\label{Eq. Birkhof matrix}
\end{eqnarray}
with: $a_0=1$ and $b_0=0$.
By combining these expressions the original set of equations can be turned into a transformed system, which is called the {\em canonical} form or Birkhoff-transform:
\begin{equation}
z\frac{d}{dz} \mathbf{F}= \mathbb{P} \mathbf{F}
\end{equation}
The coefficients of this canonical system are related to the original coefficients by the
following matrix transformation:
\begin{equation}
\frac{1}{z}\mathbbm{a} \mathbb{P} =\mathbbm{p} \mathbbm{a} - 
\frac{d}{dz} {\mathbbm{a}} 
\end{equation}
Now according to the Birkhoff theorem the $P_{ij}$ coefficients of the transformed equation will be polynomials of a degree that does not exceed the rank of the original equation. They can thus be easily obtained from the previous equation by collecting the terms in $1/z^k$ with $k=0,1$. The results are:
\begin{eqnarray}
P_{11}(z) &=& E+g^2-gz \nonumber \\
P_{12}(z) &=& -\Delta-2gb_1 \nonumber \\
P_{21}(z) &=& P_{12}(-z) \nonumber \\
P_{22}(z) &=& P_{11} (-z)
\end{eqnarray}
These are indeed polynomials of rank not higher than 1.
Note that these terms contain the expansion coefficient $b_1$ as a kind of gauge 
potential, which will have to be fixed by the quantization conditions.
The canonical form of the Rabi equation thus reads:
\begin{eqnarray}
z\frac{d}{dz} F_1 &=& (E+g^2-gz) F_1 + (-\Delta-2gb_1) F_2 \nonumber \\
z\frac{d}{dz} F_2 &=& (-\Delta-2gb_1) F_1+(E+g^2+gz) F_2
\end{eqnarray}
The original and transformed system share the same reflection symmetry:
\begin{equation}
 \left( \begin{array}{*{1}{c}} F_1(z) \\ F_2(z) \end{array}  \right)
= \pm \left( \begin{array}{*{1}{c}} F_2(-z) \\ F_1(-z) \end{array}  \right)
\end{equation}
where the plus and minus sign denote symmetric and anti-symmetric solutions respectively.
By eliminating $F_2$ the canonical set may be transformed into a second-order differential equation in $F_1$:
\begin{equation}
z^2F_1''(z) + z[1-2(E+g^2)]F_1'(z)+
[(E+g^2)- A^2 +gz-g^2z^2]F_1=0
\end{equation}
Here we have introduced $A$ to denote the gauge factor:
\begin{equation}
A=\Delta + 2gb_1
\end{equation}
By applying the Frobenius method we obtain as the roots of the indicial equation:
\begin{equation}
\rho_{\pm} = E+g^2 \pm A
\end{equation}
The differential equation can be reduced to the Kummer equation, which is solved by the confluent hypergeometric functions $_1F_1(a,b;z)$. The general solution reads:
\begin{equation}
\begin{split} F_1(z) =  C_1 &\exp(gz) \, _1F_1(1+A,1+2A;-2gz)\,z^{E+g^2 + A} \\
&+C_2 \exp(gz) \, _1F_1(1-A,1-2A;-2gz)\,z^{E+g^2 - A} \end{split}
\end{equation}
While this solution for $|z| \rightarrow \infty$ belongs to the Bargmann-Fock space, the function can only be single-valued if it is entire, this means that at least one of the roots of the indicial equation should be a non-negative integer. This provides an additional quantum condition which allows to fix the gauge factor and determines the spectrum of the Rabi Hamiltonian.
\section{Quantization of indicial roots}
The solutions of the Rabi Hamiltonian can be classified on the basis of the roots of the indicial equation. The physical requirement that the solution should belong to the Bargmann-Fock space implies the simple quantization condition that one of the roots of the indicial equation should be a non-negative integer. This criterion gives rise to different classes of solutions:
\begin{enumerate}
\item If $E+g^2$ is neither an integer nor a half integer, then only one of the roots can be integer. Indeed, suppose that both roots are integer, then one has:
\begin{equation}
\rho_+ + \rho_- = 2(E+g^2) \in \mathbb{Z}
\end{equation}
which is contrary to the starting assumption.
Hence in this case the solution of the second-order differential equation will be one-dimensional, i.e. either $C_1$ or $C_2$ must be zero, depending on which of the roots is taken to be a non-negative integer.
\item
If $E+g^2$ is half-integer, then $A$ also must be half-integer. Nonetheless the solution still remains one-dimensional, since one of the functional parameters, $1+2A$ or $1-2A$, is a negative integer or zero, and the corresponding Kummer function is not defined. 
\item
If $E+g^2$ is integer and the gauge factor $A=0$, the two first-order differential equations are uncoupled and the corresponding eigenspace is two-dimensional. 
These correspond to the Juddian exact solutions \cite{Judd, Kus} where the symmetric and anti-symmetric states cross, as was explained previously. 
The $b_1$ coefficient in this case is given by:
\begin{equation}
b_1 = -\frac{\Delta}{2g}
\end{equation} 
Note that for the solutions to be entire $E+g^2$ must be non-negative, hence the lowest Juddian baseline must have 
$E+g^2=0$.
\item
Finally it is possible that both $E+g^2$ \'and $A$ are both taken to be integer. In this case at least one of the solutions will be analytic. 
\end{enumerate}

Except for the sporadic crossing points, the solutions of the Rabi Hamiltonian will belong to the first class. In this case the requirement that one of the roots of the indicial equation should be a non-negative integer number is of paramount importance since it introduces a simple quantum number to characterize the solution space. In fact there are two
possibilities:
\begin{equation}
\rho_+ = 0,1,2,...
\end{equation}
or:
\begin{equation}
\rho_- = 0,1,2,...
\end{equation}
These two possibilities distinguish between symmetric and anti-symmetric solutions.
This can be shown as follows. Take $\rho_+=k$, with $k=0,1,2,...$. Then as explained before $\rho_-$ cannot be an integer, and the solution is one-dimensional of the following type:
\begin{equation}
 F_1(z) =  \exp(gz) \, _1F_1(1+A,1+2A;-2gz)\,z^{E+g^2 + A}
\end{equation}
By inserting this into the set of differential equations we may obtain $F_2(z)$:
\begin{equation}
 F_2(z) = \frac{1}{A} (1-2gz)F_1(z)  -\frac{1+A}{A} \exp(gz) \, _1F_1(2+A,1+2A;-2gz)\,z^{E+g^2 + A}
\end{equation}
Using the appropriate recursion formulas for the hypergeometric functions, one can easily demonstrate:
\begin{equation}
F_2(z) = (-1)^{k+1} F_1(-z)
\end{equation}
Hence if $k$ is even, the quantization of $\rho_+$ will lead to anti-symmetric solutions,
while odd $k$ values will generate the symmetric solutions.
On the other hand imposing the quantization condition for  $\rho_-$  leads to
the opposite rule, since in this case:
\begin{equation}
F_2(z) = (-1)^{k} F_1(-z)
\end{equation}
\section{Results}
%
\subsection{Recurrence and series expansion}
In addition to the quantization of $\rho$ also the expansion coefficients 
in the Laurent series are constrained by the requirement that the solution of the
initial system should belong to the Bargmann-Fock space. The recurrence relationships which determine these coefficients are more involved than in Braak's case since we now have two interrelated recurrence relations. The series expansion of Eq.\eqref{Eq. Birkhof matrix} for the $a_{11}$ matrix element yields expressions for the $a$-coefficients, from $n=1$ onwards:
\begin{equation}
a_n= \frac{1}{n}\left[-\left(\Delta + 2gb_1\right)b_n + (-1)^n\Delta\sum\limits_{\nu=1}^n g^{n-\nu}b_{\nu}+ (E+g^2)
\sum\limits_{\nu=1}^n (-1)^{\nu+1} g^\nu a_{n-\nu}\right] 
\end{equation}
By combining $na_n+ga_{n-1}$ one then obtains the first four-term recursion relation, which
generates $a_n$ form $a_{n-1}$ and $b_n$, $b_n-1$ coefficients:
\begin{equation}
na_n = (E+g^2-n+1)ga_{n-1} -\left\{ 2gb_1+\Delta[1+(-1)^{n+1}]\right\} b_n - 
g(\Delta +2gb_1)b_{n-1}
\end{equation}
Starting from $n=2$ the $b_n$ coefficients are given by: 
\begin{eqnarray}
2g b_n &= & \left(n-1\right)b_{n-1}+ (\Delta + 2gb_1)a_{n-1} +
 (-1)^n \Delta \sum\limits_{\nu=0}^{n-1} g^{n-1-\nu}a_\nu  
\nonumber \\
& &+ (E+g^2)\sum\limits_{\nu=1}^{n-2} (-1)^{n-1 +\nu} g^{n-1-\nu} b_\nu 
\end{eqnarray}
Again by combining $nb_n+gb_{n-1}$ one obtains the second five-term recursion relation, which
generates $b_n$ from the previous $a_{n-1}$, $a_{n-2}$, $b_{n-1}$, and $b_{n-2}$ coefficients.
\begin{equation}
\begin{split}
2gb_n=& g (\Delta + 2gb_1)a_{n-2} + \left\{ 2gb_1
 +\Delta[1+(-1)^n] \right\}a_{n-1} \\
&+g\left[ n-2 -(E +g^2)\right] b_{n-2} + (n-1-2g^2) b_{n-1}
\end{split}
\end{equation}
Consecutive coefficients are thus obtained by leapfrogging the recurrence relations from $b_1$ onwards: 

\noindent $b_1 \rightarrow a_1 \rightarrow b_{2} \rightarrow a_{2} \rightarrow b_3 \rightarrow ..$. 

\subsection{The eigenvalue equation}
Everything is now in place to obtain the eigenvalue equation.
Again it is based on limiting the recurrent series at either end. At the lower end of the series the zeroth-order parameters are fixed by the limiting unit matrix, and the first-order $b_1$ coefficient is quantized by the requirement on the $\rho$ parameter. 
One thus chooses a value of $k$ and the plus or minus root of the indicial equation.
This is sufficient to start the recurrence:
\begin{eqnarray}
a_0 &=& 1 \nonumber \\
b_0 &=& 0 \nonumber \\
b_1 &=& \frac{k- (E+g^2) \mp \Delta}{2g}
\end{eqnarray}
From this input onwards all higher-order coefficients are generated by the leapfrog recurrence relations.
For a fixed integer $k$ and a choice of the parity sign of the roots, the only remaining variable in which all coefficients will be expressed is the energy.
To determine then the energy we follow the same argument as before and require that the
series terminate at the high end, i.e. we require:
\begin{equation}
\lim_{n\rightarrow \infty} b_n(E+g^2) = 0
\end{equation}
Alternatively one could also terminate the series at $a_n$, yielding the same results. The $b_n$ or $a_n$ coefficient is a polynomial in $E+g^2$, the roots of which again will determine the eigenvalues. This is a new kind of transcendental function. It is determined by three choices: $\rho_+$ versus $\rho_-$, the value of $k$, and the value of $n$.  Concerning this last choice, in Figs. 4 and 5 we plot the functions $b_n(E+g^2)$ for $n=5,6,7,8$, and $k=0,1$ for anti-symmetric and symmetric roots respectively. These functions clearly resemble Kummer-type polynomials themselves \cite{Slater}. While the actual numerical calculations are quite time-consuming for larger $n$ it is remarkable that even for 
small $n$ values the previous results of Moroz \cite{Moroz} and Braak \cite{Braak} are easily recovered.
However truncating the series at higher $n$-value will lead to sharper warping of the functions and hence more accurate eigenvalues. For convenience the lower eigenvalues are listed in Table 1.
\begin{table}
\caption{Eigenenergies $E+g^2$ for $\Delta=0.4$ and $g=0.7$ ($b_9=0$).}
\begin{tabular}{l|l}
			Symmetric	& Anti-symmetric \\
\hline
$+0.062956$ & $-0.217805$   \\
$+1.163604$ &  $+0.86095$   \\
$+1.85076$ & $+2.12701$   \\
$+3.03523$ &  $+2.9567$   \\
$+4.0569$ &  $+3.95113$ \\
 \end{tabular}
	\label{Table1}
\end{table}
\begin{figure}
        \includegraphics*[width=12cm]{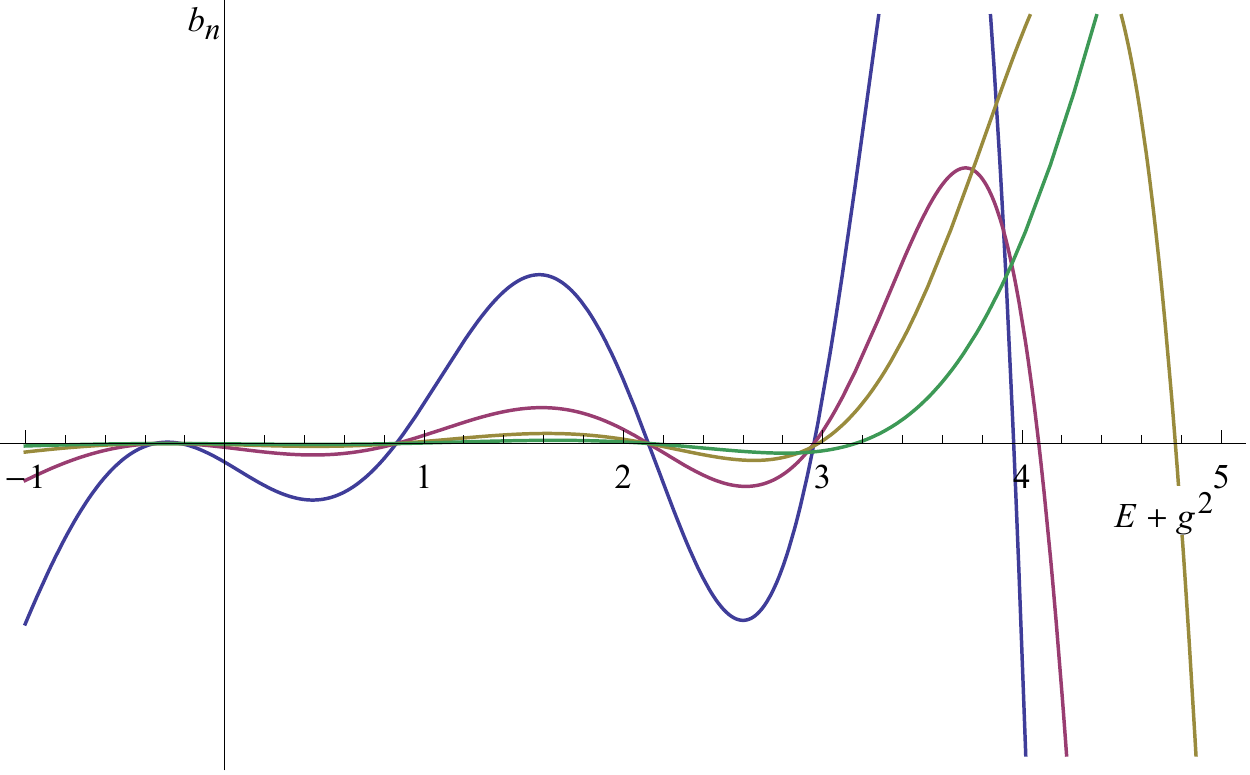}
        \caption{Plot of the transcendental function $b_n(E+g^2)$ for $n=5$ (green), $n=6$ (brown), $n=7$ (purple), $n=8$ (blue). The $k$-value is set to zero, and the $\rho_+$ root is chosen. The roots correspond to the anti-symmetric eigenvalues.}
        \label{fig4}
\end{figure}
\begin{figure}
        \includegraphics*[width=12cm]{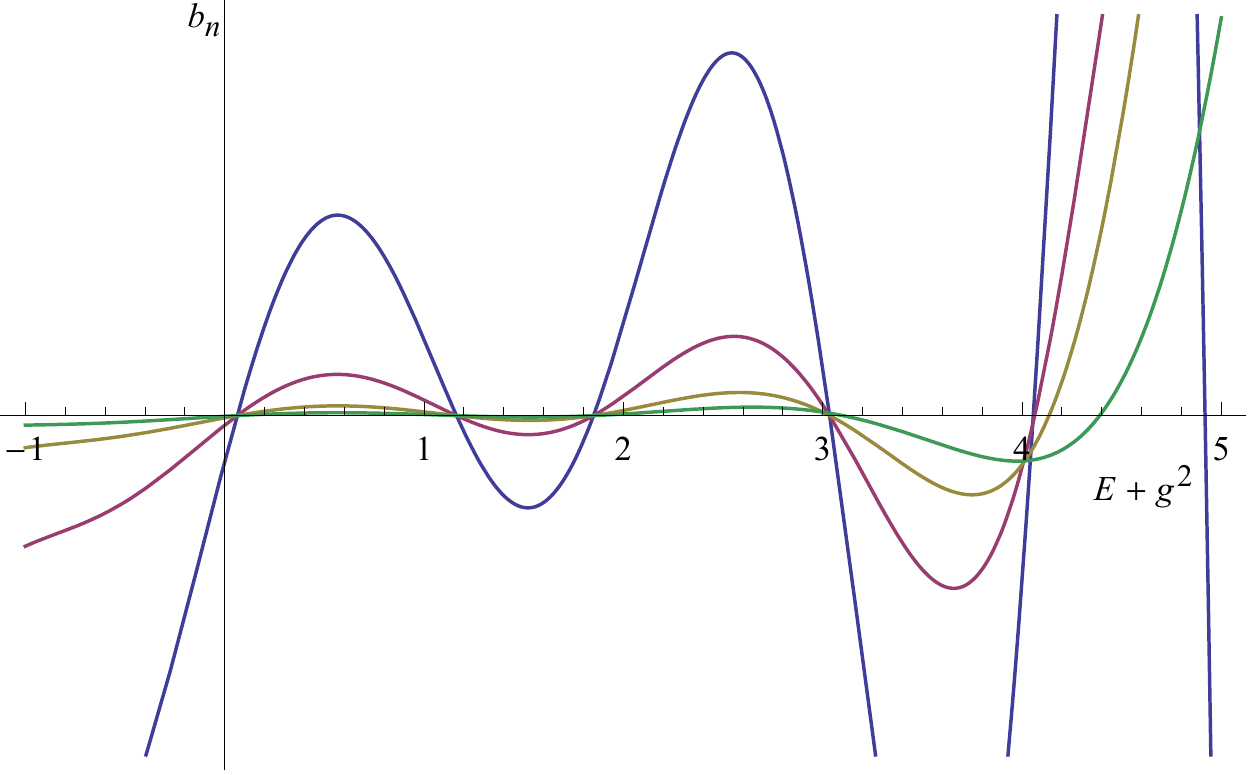}
        \caption{Plot of the transcendental function $b_n(E+g^2)$ for $n=5$ (green), $n=6$ (brown), $n=7$ (purple), $n=8$ (blue). The $k$-value is set to one, and the $\rho_+$ root is chosen. The roots correspond to the symmetric eigenvalues.}
        \label{fig5}
\end{figure}
In the next two figures we take $b_8=0$, and $\rho_+=k$, with even (Fig. 6) and odd (Fig. 7) values for $k$; in this way one recovers the anti-symmetric and symmetric eigenenergies respectively. Most importantly as the figures show, different $k$ values reproduce the same spectrum, but with higher $k$-values the higher energy roots can be obtained more accurately. In Fig. 6 some additional roots appear at the low-energy end of the spectrum which are non-physical. We will return to this point in the next section.
\begin{figure}
        \includegraphics*[width=12cm]{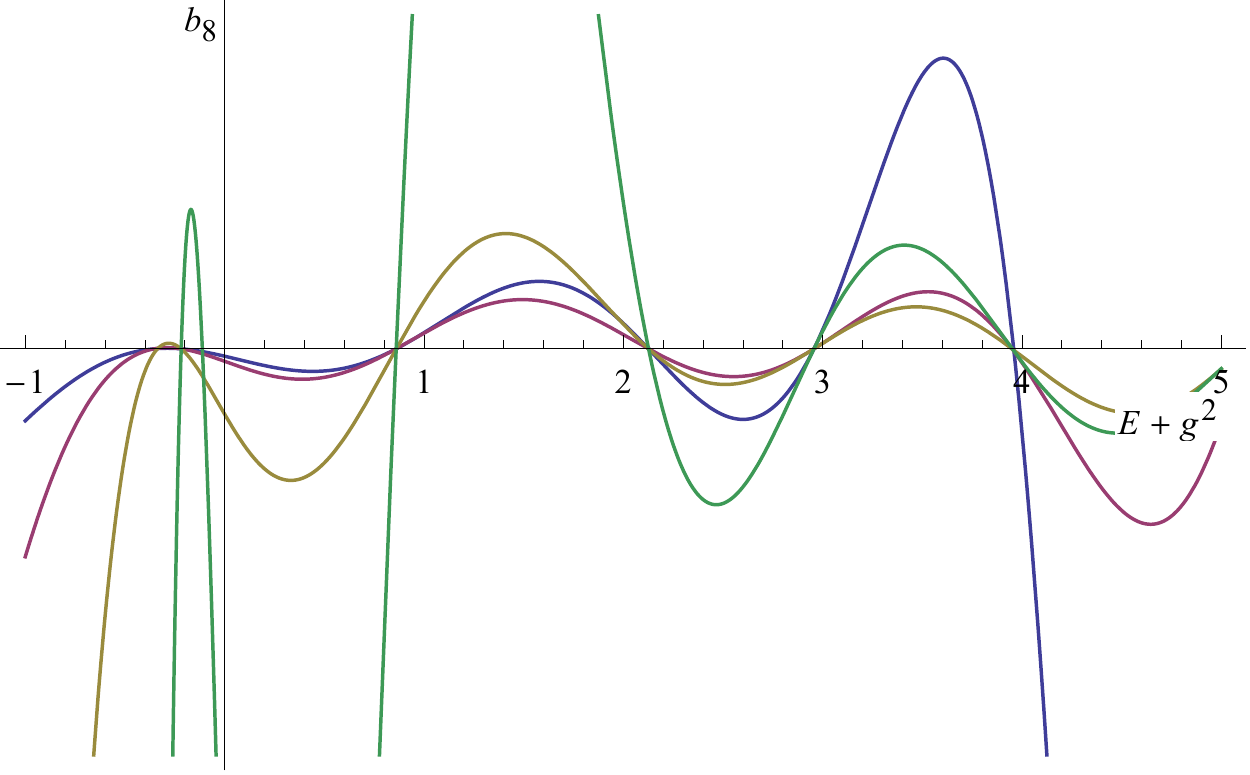}
        \caption{Plot of the transcendental function $b_n(E+g^2)$ for $n=8$,
based on the expression for $\rho_+$  with different even $k$-values: $k=0$ (blue), $k=2$ (purple),  $k=4$  (brown), $k=6$ (green).
The roots correspond to eigenenergies of anti-symmetric states.} 
        \label{fig6}
\end{figure}
\begin{figure}
        \includegraphics*[width=12cm]{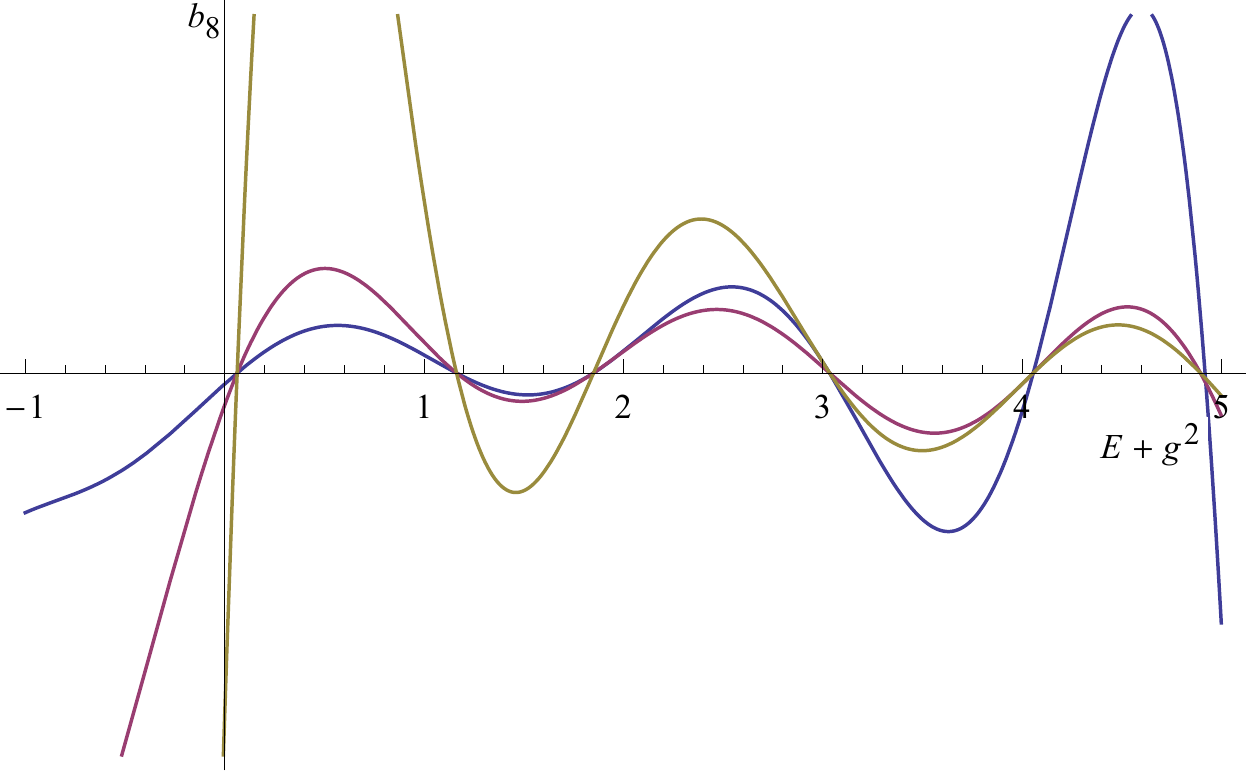}
        \caption{Plot of the transcendental function $b_n(E+g^2)$ for $n=8$,
based on the expression for $\rho_+$  with different odd $k$-values: $k=1$ (blue), $k=3$ (purple),  $k=5$  (brown).
The roots correspond to eigenenergies of symmetric states.} 
        \label{fig7}
\end{figure}  
\section{Discussion}
\subsection{The quantum number $k$}
When solving the Birkhoff equation an integer number $k$ naturally appears which we may associate with the quantum number that was used by Braak to label states of the same reflection symmetry. Let us study this number into more detail.
It relates to the energy as:
\begin{equation}
k= E + g^2 \pm A
\end{equation}
The gauge parameter $A$ is seen to measure the distance between the actual eigenenergy and the Juddian baselines, $E_k$. These baselines are straight decreasing lines in an $\{E,g^2\}$ plot, defined by:
\begin{equation}
E_k = k-g^2
\end{equation}
In the limit of strong coupling, the surface consists of two deep wells with the same energy, and hence a nearly degenerate boson spectrum, which in the limit coincides with the Juddian baselines. Reflection symmetry adaptation yields for each baseline one symmetric and one anti-symmetric state. Above this limit the appearance of the $A$ parameter makes it possible to associate a given eigenenergy with a Juddian baseline. So in a sense the baselines acts as 'attractors' for the eigenenergy. However this association is not uniquely defined, since as was shown in Figs. 5 and 6, different $k$-values can lead to the same eigenenergies. For the association to be more strict one should have a closer look at the eigenfunctions themselves.
It seems always to be the case that the $a_n$ and $b_n$ coefficients show the fastest convergence when the $k$ value corresponds to index of the nearby Juddian baseline. Increasing the value of $k$ for the same eigenenergy will simply increase $A$ by the same integer, and thus increase the rank of the Kummer function \cite{Slater}. Since the eigenfunction should not change, this increase is countered by a concomitant shift in the $a_n$ and $b_n$ coefficients, putting more weights on coefficients with larger $n$ and thus reducing the rate of convergence. The fact that different values of $k$ can still give rise to the same overall eigenfunctions is a result from the 'contiguity' relations of the confluent hypergeometric functions.
In short a given eigenstate should always be labeled by the quantum number $k$ which corresponds to the smallest distance $A$ between the actual eigenenergy and the $E_k$ Juddian baseline.

When $A$ becomes zero the level crosses the baseline, and as the solution of the Birkhoff canonical equation shows, the $\rho_+$ and $\rho_-$ roots will then coincide, implying that both the anti-symmetric and symmetric states cross simultaneously. Moreover both solutions will approach the baseline from opposite sides since their $A$-factors differ in sign. As a result the number of eigenstates in between two baselines will be constant: if one eigenstate leaves this region another one will enter at the same time. The number of states between two baselines can easily be determined from the zeroth-order spectrum at $g=0$.
For half integer values of $A$, one of the parameters $1\pm2A$ in the Kummer function will be a zero or a negative integer, and the function will not converge. Hence the values $A=\pm1/2$ in between the Juddian baselines define a 'separatrix' in between the Juddian baselines.  
\subsection{Existence}
Quantum mechanics requires that the eigenfunctions should belong to the Bargmann-Fock space, which is the set of all entire functions with a finite norm. This criterion should apply to the solutions $f_1(z)$ and $f_2(z)$ of the original Hamiltonian equations. It certainly applies to the $F_1(z)$ and $F_2(z)$ solutions of the canonical equations. The Kummer series $_1F_1(a,b;z)$ is absolutely convergent for all values of the parameters, except for  $b=0,-1,-2,...$ where it has simple poles. It is furthermore single-valued and differentiable for all values of $z$, real or complex. For $a=-n$, with $n=0,-1,-2,...$, it becomes a finite polynomial of degree $n$. With $k$ a non-negative integer the solutions are moreover entire.
However the fact that the solutions of the canonical equations are within the Bargmann-Fock space, does not necessarily imply that the actual eigenfunctions of the original equation are too. Indeed it is found that for the low-energy end of the spectrum one may obtain roots of the transcendental equation which are non-physical. A case in point is the $k=6$ polynomial for $b_8=0$ in Fig. 6, which has a non-physical root at -0.111577.  This could point to numerical inaccuracies, as it disappears for $b_9=0$, and requires further investigation.
%
\section{Conclusions}
By transforming the Rabi equations to their canonical form an interesting new perspective is opened on this simplest case of boson - two-state-fermion coupling.
The canonical equations introduce a gauge potential which measures the distance between the actual eigenenergy and the integer quantum numbers of the Juddian baselines. In this way they draw attention to the controlling role of these baselines
on the spectrum. An important result from the treatment is that the number of levels in between two baselines is constant, and can easily be determined from the zeroth-order spectrum.  As we have argued in a previous paper \cite{Ceulemans1997} the canonical equation for the $E\times e$ Jahn-Teller Hamiltonian is identical to the Rabi case. In a forthcoming publication we will report the results of an analogous treatment on the Jahn-Teller case.

\bibliography{Rabi}
\clearpage

\end{document}